\documentclass{kluwer}
\newdisplay{guess}{Conjecture}
\usepackage{epsfig}
\begin{document}                                                                                   
\begin{article}
\begin{opening}         
\title{The Recurrent Variational Approach applied to the Electronic
Structure of Conjugated Polymers} 
\author{St\'ephane \surname{Pleutin}$^{1}$, Eric
\surname{Jeckelmann}$^{2}$, Miguel A. \surname{Mart\'{\i}n-Delgado}$^{3}$ and German \surname{Sierra}$^{4}$}  
\runningauthor{S. Pleutin, E. Jeckelmann, M.A Mart\'{\i}n-Delgado, G. Sierra}
\runningtitle{The RVA applied to the Electronic
Structure of Conjugated Polymers}
\institute{$^{1}$ Max-Planck-Institut f\"ur Physik Komplexer Systeme,
Dresden, Germany\\
$^{2}$ Fachbereich Physik, Philipps-Universit\"at Marburg, Marburg, Germany\\
$^{3}$ Departamento de F\'{\i}sica Te\'orica I, Univers\'{\i}dad
Complutense, Madrid, Spain\\
$^{4}$ Instituto de Matem\'aticas y F\'{\i}sica Fundamental, C.S.I.C.,
Madrid, Spain}
\date{July 12, 1999}

\begin{abstract}
In this note, first the Recurrent Variational Approach (RVA) is
introduced by using as example a non--trivial spin--model, the
spin--1/2 antiferromagnetic two--leg--ladder. Then, a first
application of this scheme to the electronic structure of conjugated
polymers is proposed. An analogy between the two--leg ladder and the
dimerized chain described by the Extended Peierls--Hubbard Hamiltonian
is underlined and comparisons are made with DMRG calculations. The
proposed ansatz is useful to get some analytical insight and it is a good
starting point for further improvements using the RVA.
\end{abstract}
\end{opening}

\section{Introduction}
Conjugated molecules have attracted the attention of chemists and physicists
since the early days of quantum mechanics. A remarkable example is
given by H\"uckel who developed
in the thirties the first independent--electron theory applicable to
polyenes \cite{huckel}. In the sixties, J.A. Pople and
S.H. Walmsley \cite{pople}, introduced by using this simple theory in their
pioneering work what would become twenty years later the solitonic
excitations. Following this
line, Su, Schrieffer, Heeger and, independently, Rice, proposed a
model specifically devoted to the polyacetylene -- known as the SSH
model -- where the role of the
electron--phonon interaction is especially emphasized \cite{ssh,rice};
here, the polyacetylene is thought of as a Peierls insulator and the
solitons, polarons and bipolarons as the major relevant excitations in
this system. A very nice review of these early studies is included in
the book of L. Salem \cite{salem} and a review about the SSH model
can be found in the next reference \cite{heeger}.

As a counterpart to this one--electron theory, the major role of electron
correlation was soon recognized. In the fifties, the celebrated
Pariser--Parr--Pople (PPP) Hamiltonian was suggested as a possible
candidate for the studies of conjugated oligomers \cite{PP,P}. Later,
Ovchinnikov et al. suggested to consider conjugated polymers as
Mott--Hubbard instead of Peierls insulators \cite{ovchinnikov}. Then,
one has to face a N--body problem with a strong Coulomb interaction
with a non--zero range;
it was subject to many numerical works such as exact diagonalizations of
very small oligomers \cite{chandross}, selected CI calculations
\cite{tavan} and, quite recently, Density Matrix Renormalization
Group studies \cite{fano,bursill,eric,shuai}. The PPP Hamiltonian was also a preferential model to
support the development of new methods before applying them to more
realistic models; a very good example of such work is given by the
Coupled--Cluster method first implemented in quantum chemistry for
the PPP Hamiltonian \cite{cizek}.

Today, it is well recognized that both the
electron--phonon and electron--electron interactions are equally
important \cite{revue}. Moreover, interchain interaction and disorder are also of
importance. Therefore, the problems one has to face to fully understand
conjugated polymers stay quite challenging as it is the case for most of
the recent organic compounds including high Tc
superconductors. Despite a large amount of work, from the early
thirties to now, the main questions
remain awaiting an answer! For instance, a comprehensive picture
for the ground state of these systems is still needed; it is the goal
of this work to propose a relatively simple and, we believe,
promising, way to understand the ground state of conjugated polymers.

More generally, one dimensional correlated electron systems appear
over the years as a growing and important field in condensed matter
theory. One main reason of that comes from the fact that these
systems serve as a theoretical laboratory to explore new methods of solution,
analytically or numerically. A second main reason
-- we would say more physical -- comes from the fact that over the years,
more and more experimental realisations of these systems appear. New
concepts emerge from such studies as the ones included in the phenomenology of
Luttinger--liquids \cite{haldane}.

Among the technical methods proper to the one dimensional geometry,
one may cite the Bethe ansatz \cite{bethe}, the bosonization
techniques \cite{haldane}, and, more recently, the Density--Matrix
Renormalization Group (DMRG) method \cite{white,dmrg} and a closely related scheme which
is directly considered in this note, the Recurrent
Variational Approach (RVA) \cite{RVA1,dmrg}. The two first methods are analytical and
the third one is numerical; the RVA method is in
between.

The RVA, presented here, is a variational method invented recently in
the context of spin--ladders \cite{RVA1}. It is a method very closely
related to the DMRG scheme with the main differences coming from the
fact that only one state is retained as the best candidate for the
ground state in the RVA on the contrary to the DMRG which considered
much more states \cite{dmrg}. Most often, the results obtained with it are less accurate
than the DMRG ones, but it is much easier to get a physical insight into the problem; the analysis of the results is simpler and the RVA could provide a "physical" picture of some phenomena which,
we believe, are of utmost importance in the field of conjugated polymers
where the most important ingredient are still not fully recognized
\cite{excitations}.

This note is organized as follows. In section II, the RVA is presented
using the example of the two--leg spin ladders. In section III, the
RVA method is applied to the study of the dimerized chain described by
the Extended Peierls--Hubbard model. A nice similarity between the
two--leg ladder and this system is pointed out and
comparisons with DMRG calculations are made. We conclude in section IV
and give some reasonable perspectives.

\section{RVA method applied to two--leg Spin Ladders}

A spin ladder is an array of coupled spin chains. The horizontal
chains are called the legs, the vertical ones, rungs. In the case of
spin one--half antiferromagnet spin--ladders, these systems show a
remarkable behaviour in function of the number of leg: there is a gap
in the excitation spectrum of even--leg
ladders and, on the contrary, no gap in the excitation spectrum
of odd--leg ladders. In terms of correlation lengths, this means that
there is short (long) --range spin correlation in even (odd) --leg
ladder (see \cite{ladder} for a review). 

The even--leg ladders show a spin-liquid ground state described by a
Resonance Valence Bond (RVB) state introduced by P.W. Anderson
\protect\cite{anderson}; such systems may be efficiently described by
using a direct--space method as the DMRG and RVA ones. In this note, we
will consider only the simplest even--leg ladder, the two--leg ladder, and we
will describe it with the simplest RVB state, the Dimer--RVB state
where the elementary singlets are between two nearest--neighbour
sites. This problem is easily solved by using the RVA method
\cite{RVA1}. Moreover, with the same scheme, it is possible
straightforwardly to go beyond the simple
Dimer--RVB state by considering more extended elementary singlets \cite{RVA1}, and to study more complex systems as
the four--leg ladders for instance \cite{RVA4}. In principle, this
method should be relevant for every one--dimensional system with
exponentially decreasing correlation lengths as conjugated polymers
\cite{pleutin2}.

\begin{figure}[h]
\centerline{\epsfig{file=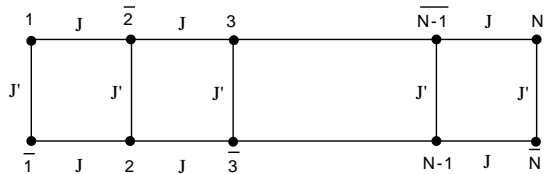,width=17pc}}
\caption[]{The two--leg ladder with an AF--Heisenberg model described by two 
coupling constants: $J$ for the links in the legs, 
and $J'$ for the vertical rungs}
\label{ladder}
\end{figure}

In this
note we consider the simplest case given by the two--leg ladder
(see figure (\ref{ladder})). The corresponding Hamiltonian is the
following 
\begin{equation}
H = J  \sum_{k=1}^{N-1}
( {\bf S}_k \cdot {\bf S}_{\overline{k+1}} + 
{\bf S}_{k+1} \cdot {\bf S}_{\overline{k}} ) + 
J'  \sum_{k=1}^N {\bf S}_k \cdot {\bf S}_{\overline{k}}
\label{heisenberg}
\end{equation}where $J$ ($J'$) are the exchange integral for the leg
(rung) with $J>0$ ($J'>0$). $N$ denotes the number of rungs of the
ladder and open periodic boundary conditions have been assumed.

\begin{figure}[h]
\centerline{\epsfig{file=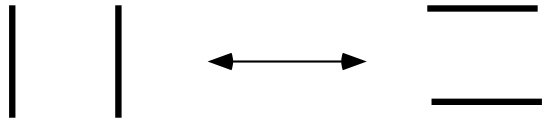,width=17pc}}
\caption[]{The basic bond resonance mechanism between horizontal and 
vertical bonds in an elementary plaquette of four sites.}
\label{resonance}
\end{figure}

In the limit where $J'$ is much larger than $J$, the ground state of
(\ref{heisenberg}) is simply given by the product of singlets localized
along the rungs (see fig. (\ref{0resonon})). Starting from this limit,
one may include fluctuations around it to treat the general case
described by
the Hamiltonian (\ref{heisenberg}). A minimal way to do it, is to
consider instead of singlet on rungs, pairs of nearest neighbour
singlets on the leg; these local fluctuations are called resonon in the
rest of the paper. Therefore, the resonance mechanism -- in the sense of Pauling
\cite{pauling} -- we will consider in the following are represented in
figure (\ref{resonance}).

With the two selected local configurations, the singlet localized on
a rung and the resonon, we have to generate the corresponding Hilbert
space; the configurations are characterized by the number of resonon, $M$,
and their positions, $\{ x_i\}$, as it is shown in the next
examples. The Hilbert space is splitted according to the number of
resonon; we have then to consider


\begin{itemize}
\item the zero--resonon sector with one unique state $|0\rangle$ (Fig.~\ref{0resonon}),
\begin{figure}[h]
\centerline{\epsfig{file=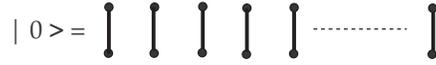,width=15pc}}
\caption[]{The zero--resonon state $|0\rangle$ 
in a two--leg ladder. It is made of all vertical 
rung singlets. It is the reference state in the 
strong coupling limit $J' \gg J$.}
\label{0resonon}
\end{figure}

\item the one--resonon sector with the states $\{ |x_1\rangle \}$ (Fig.~\ref{1resonon}),
\begin{figure}[h]
\centerline{\epsfig{file=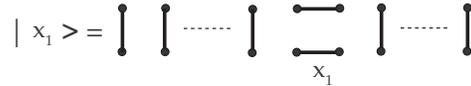,width=16pc}}
\caption[]{A generic one--resonon state $|x_1\rangle$ 
in a two--leg ladder.}
\label{1resonon}
\end{figure}

\item the two--resonon sector with the states $\{ |x_1,x_2\rangle \}$ (Fig.~\ref{2resonon}),

\begin{figure}[h]
\centerline{\epsfig{file=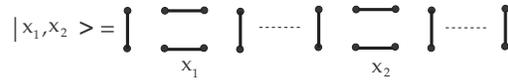,width=17pc}}
\caption[]{A generic two--resonon state $|x_1, x_2\rangle$ 
in a two--leg ladder.}
\label{2resonon}
\end{figure}

\noindent etc...

\item the ${N\over 2}$--resonon sector with one unique state $|x_1,x_2,\ldots,x_{M}\rangle$
(Fig.~\ref{N/2resonon}).

\begin{figure}[h]
\centerline{\epsfig{file=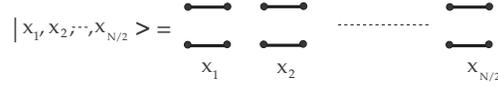,width=17pc}}
\caption[]{The all--resonon state $|x_1, x_2,\ldots, 
x_{{N\over 2}}\rangle$ in a two--leg ladder.}
\label{N/2resonon}
\end{figure}

\end{itemize}

With this classification of states we may write the following generalized 
Dimer--RVB state for a two--leg spin ladder,

\begin{equation}
|\psi(u)\rangle_{RVB} = \sum_{M=0}^{[N/2]} u^M 
\sum_{\{ x_1,x_2,\ldots,x_{{M}}\}} |x_1,x_2,\ldots,x_{M}\rangle
\label{18}
\end{equation}

\noindent where the sum is taken over all sectors with $M$ resonons
and $u$ is the amplitude for a  resonon.
Here $u$ is a  variational parameter to be determined 
upon minimization of the ground--state energy. 
As a result, it is a function of the ratio of couplings, namely,
$u=u(J/J')$.

Since we kept only two very localized building blocks, it is possible
to generate the state (\ref{18}) in a recursive manner which we
present now.

\subsection{Recurrence relations for the wave function}
Denoting by $|N\rangle \equiv |\psi(u)\rangle_{RVB} $ a
generic RVB state (\ref{18}) for a two--leg ladder of $N$ rungs, there
are only two possibilities or {\it movements} to create generic RVB
states of higher length, namely,

\begin{itemize}

\item addition of  one vertical rung to create the state $|N+1\rangle$,

\item addition of  one pair of horizontal bonds (resonon) to create 
$|N+2\rangle$.
  
\end{itemize}

From these arguments we can establish that the Dimer--RVB states (\ref{18})
satisfy a {\it recursion relation} given by,

\begin{equation}
|N+2\rangle = |N+1\rangle \otimes |\phi_1\rangle_{N+2} + 
u \; |N\rangle \otimes |\phi_2\rangle_{N+1,N+2} 
\label{23}
\end{equation}

\noindent where the state  denoted by $|\phi_1\rangle_{N+2}$
is a vertical rung at position $N+2$
and
$ |\phi_2 \rangle_{N+1,N+2}$ 
is made up of a pair of horizontal  
bonds located between the rungs at $(N+1,N+2)$, i.e.,

\begin{equation}
|\phi_2 \rangle_{N+1,N+2} = 
(N+1,\overline{N+2})  (N+2,\overline{N+1}) 
\label{24}
\end{equation}

\noindent where $(N,\overline{M})$ denotes a singlet between the
sites $N$ and $\overline{M}$.   

Using (\ref{23}) one can generate recursively the 
Dimer RVB state $|N\rangle$ from previous states with lower 
length and  estimate variationally the ground--state energy as,

\begin{equation}
e^{(g.s.)}_{N} \equiv {\langle N|H_{AFM}^{{ two-leg}}|N\rangle \over
(2N) \langle N|N\rangle } = e(N,J'/J;u)
\label{25}
\end{equation}

\noindent The value of $u$ is fixed by minimization of (\ref{25}).

\subsection{Recurrence relation for the norm}

To compute  the norms of the states (\ref{18}) let us define

\begin{equation}
Z_N = \langle N|N\rangle 
\label{26}
\end{equation}

\noindent The two selected local configurations are not orthogonal, it
is then convenient to define an auxiliary function $Y_N$ as follows,

\begin{equation}
Y_N =  \big(_{N}\!\langle \phi_1| \otimes \langle N -1|\big)| N \rangle
\label{27}
\end{equation}

\noindent The second order RR's for the states leads 
(\ref{23}) to a closed set of RR's for the overlaps $Z_N ,Y_N $ , namely,

\begin{equation} 
 \begin{array}{rl}
Z_{N+2} = &  Z_{N+1} + u Y_{N+1} + u^2 Z_N \\
Y_{N+2} = &  Z_{N+1} + {u\over 2} Y_{N+1} \end{array}
\label{28} 
\end{equation}

\noindent where we have made use of the result,

\begin{equation}
_{N+1}\langle \phi_1 | \phi_2 \rangle_{N,N+1} = {1\over 2} |\phi_1 \rangle_N
 \label{29}
\end{equation}

The RR's (\ref{28}) together with the initial conditions,

\begin{equation}
\begin{array}{ll} 
Z_{0} = Z_{1} = 1 & \; ,Y_{0} = 0, \  Y_{1} = 1 \end{array}
\label{30}
\end{equation}

\noindent determine 
$Z_N$ and $Y_N$ for arbitrary values of $N$ as functions of 
$u$. 

Note that instead of introducing the function $Y_N$, it is also
possible to make first linear combination of our building blocks, making used of (\ref{29}) \cite{eder}, in
order to guaranty the orthogonality between the selected local
configurations. The overlap
would then be given by
only one recurrent relation as it will be the case in the next section.

\subsection{Recurrence relation for the expectation value of the Hamiltonian}

The expectation values of the Hamiltonian can be determined following
the same steps outlined above. To this end,
we  introduce the quantities,

\begin{equation} 
\begin{array}{ll}
E_N = \langle N| H_N |N \rangle & \; ,D_N = \big(_{N}\!\langle \phi_1|
\otimes \langle N -1|\big) H_N |N \rangle
\end{array}
\label{31}
\end{equation}

\noindent To obtain the  RR's satisfied by  $E_N,D_N$,
one  splits the Hamiltonian 
$H_N $ of a two--leg ladder of 
length $N$ (\ref{heisenberg}) into two pieces:

\begin{equation} 
H_N = H_{N-1} + H_{N-1,N}
\label{32}
\end{equation}

\noindent where $H_{N-1}$ is the Hamiltonian of length $N-1$ and $H_{N-1,N}$
is the rest of the whole $H_N$ Hamiltonian, 
$H_{N-1,N}\equiv H_N - H_{N-1}$, which is made of one vertical rung
and two horizontal links. With this splitting and using  
(\ref{23}) and (\ref{26}), we obtain,

\begin{equation} 
\begin{array}{cl}
E_{N+2} = & E_{N+1} + J' \epsilon_0 Z_{N+1} 
  + u ( D_{N+1} + (2J+J') \epsilon_0 Y_{N+1} ) \\ 
 & + u^2 (E_N + 2J\epsilon_0 Z_N) \\
D_{N+2} = &  E_{N+1} + J' \epsilon_0 Z_{N+1} 
 + {u\over 2} ( D_{N+1} + (2J+J') \epsilon_0 Y_{N+1} ) 
\end{array}
\label{33}
\end{equation}

\noindent where $\epsilon_0 = -3/4$ is the lowest eigenvalue of the operator 
${\bf S}_1 \cdot {\bf S}_2$. 

The initial conditions for $E_N$ and $D_N$ are

\begin{equation} 
\begin{array}{ll}
E_{0} = 0, \  E_{1} = J' \epsilon_0 & D_{0} = 0, \  D_{1} = J' \epsilon_0.
\end{array}
\label{34} 
\end{equation}

\noindent The RR's for the energies involve the norms of the states
and depend both on $u$ and the coupling constants $J',J$.

By working with orthogonal building blocks \cite{eder}, it is also
possible to go back to only one recurrent equation for the expectation
value as for the norm; this will be the case in section 3.

\subsection{Results for the Variational Ground--State Energy}

The ground--state energy for a ladder of length $N$ is estimated.  In
the thermodynamic limit $N \rightarrow
\infty$ one can find a closed expression for the density energy per
site \cite{RVA1},

\begin{equation}
e_{\infty} = \lim_{N \rightarrow \infty}\frac{1}{2N} \frac{E_N}{Z_N}
= \frac{ R( \alpha) }{ 2 \alpha Q'(\alpha) P( \alpha) }
\label{49}
\end{equation}

\noindent in terms of three polynomials $P, Q, R$ evaluated at the
biggest root $\alpha$ of the cubic polynomial $Q(y)$.  Finally one
looks for the absolute minimum of $e_{\infty}$ by varying the
parameter $u$.  In Table~\ref{Tab2} we show the ground--state energies
per site for different values of the coupling constant ratio $J/J'$,
varying through strong, intermediate and weak coupling regimes.

\begin{table}
\caption{The values $-e_{\infty}^{{\rm MF}}/J'$ are mean field (MF)
values taken from \protect\cite{GRS}, while $-e_{\infty}^{{\rm
Lan}}/J'$ are Lanczos values taken from \protect\cite{Barnes}.}
\begin{tabular}{lllll} 
\hline \rule[-1.5mm]{0mm}{5mm}
$J/J'$ \ \ \ & $u$ \ \ \ \ \ \ \ \ \ \ \ \ \ \ & $-e_{\infty}^{{\rm RVA}}/J'$\ \ \ &   
$-e_{\infty}^{{\rm MF}}/J'$ \ \ \ & $-e_{\infty}^{{\rm Lan}}/J'$\\  
\hline

  $0$ & $0$ & $0.375$  & $0.375000 $  & $$ \\ 

  $0.2$ & $0.128521$ & $0.383114$  &  $0.382548$  & $$ \\

  $0.4$ & $0.323211$ & $0.40835$  &  $0.405430$  & $$ \\ 

  $0.6$ & $0.578928$ & $0.44853$  &  $0.442424$  & $$ \\ 

  $0.8$ & $0.87441$ & $0.499295$  &  $0.489552$  & $$ \\ 

  $1$ & $1.18798$ & $0.556958$  &  $0.542848$  & $0.578$ \\ 

  $1.25$ & $1.58519$ & $0.63518$  & $0.614473$  & $0.6687$ \\ 

  $1.66$ & $2.21853$ & $0.772172$  &$0.738360$  & $0.8333$ \\ 

  $2.5$ & $3.39153$ & $1.06915$  &$1.002856$  & $1.18$ \\ 

  $5$ & $5.9777$ & $1.99285$  & $$  & $2.265$ \\ 
\hline
\end{tabular}
\label{Tab2}
\end{table}

In the strong coupling regime $J/J' < 1 $ the RVA states
give a slightly better ground--state energy than the mean field result.
This latter state produces rather unphysical results for $J/J' >1$,
which does not occur in our case. The RVA results compare also well
with the exact results for $J' \sim J$. 

\begin{figure}[h]
\centerline{\epsfig{file=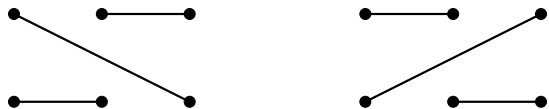,width=17pc}}
\caption[]{Local configurations extended over three rungs (the knight move)}
\label{knightmove}
\end{figure}

The RR's can also be used to compute the spin correlator $\langle {\bf
S}_i \cdot {\bf S}_j \rangle$ which has an exponential decay behaviour
${\rm exp}(-|i-j|/\xi)$, with $\xi$ the spin correlation length which
satisfies the equation

\begin{equation}
u^3 {\cal L}^3 - (2 + u) u^2 {\cal L}^2 - (2 + 4 u ) u^2 {\cal L}
+ 4 u^3 = 0
\label{xi}
\end{equation}

\noindent
where ${\cal L} = {\rm e}^{1/\xi}$. For $J=J'$ one finds $\xi = 0.737$, which can be compared with its exact value given by 3.2. The latter results imply that at the isotropic point the
bonds extended over three rungs are quite important.  The simplest improvement of the
Dimer--RVB state is to add a bond of length $\sqrt{5}$, analogue to a
knight move, shown in figure (\ref{knightmove}); this first
improvement leads to a third order RR. The ground state  energy per site of this state is given by -0.5713 while the
spin correlation length is 0.959. Hence there is an
improvement in both quantities but still one needs longer bonds.

\begin{table}
\caption{ Here $m$ is total the number of multiplets, $N_A$ is the number
of independent variational parameters, $d_J$ is the number of multiplets
with spin $J$, $e_{\infty}^{{\rm MP},{\rm DMRG}}$ is 
the g.s. energy density of the matrix--product state (DMRG), $1-P_m$ is the
probability of the states truncated out in the DMRG 
and $\xi^{\rm MP}$ is the spin correlation 
length of the MP state. The exact results
are given by $e_{\infty} = 1.4014845$ and $\xi= 6.03$ \protect\cite{white-huse}.}
\begin{tabular}{lllllllllllll}
\hline \rule[-1.5mm]{0mm}{5mm}
$m$ && $N_A$ & $d_{1/2}$ & $d_{3/2}$ & $d_{5/2}$ & $-e_\infty ^{\rm MP}$ && $%
-e_\infty ^{\rm DMRG}$ && $1-P_m$ && $\xi ^{\rm MP}$ \\ 
\hline
$1$ && $0$ & $1$ & $0$ & $0$ & $1.333333$ && $1.333333$ &&  
$1.6\times 10^{-2}$ && 0.910  \\ 
$2$ && $2$ & $1$ & $1$ & $0$ & $1.399659$ && $1.369077$ & &$1.4\times 10^{-3}$
&&2.600  \\ 
$3$ && $4$ & $2$ & $1$ & $0$ & $1.401093$ & &$1.392515$ & &$1.3\times 10^{-5}$
&&3.338  \\ 
$4$ && $7$ & $2$ & $2$ & $0$ & $1.401380$ & &$1.401380$ & &$1.6\times 10^{-5}$
&& 3.937 \\ 
$5$ && $10$ & $2$ & $2$ & $1$ & $1.401443$ & &$1.401436$ && $7.6\times 10^{-6}$
&& 4.085  \\ 
$6$ && $13$ & $2$ & $3$ & $1$ & $1.401474$ & &$1.401468$ && $1.3\times 10^{-6}$
&& 4.453\\
\hline
\end{tabular}
\label{Tab3}
\end{table}

The RVA method has already been applied to several model systems: to
the Heisenberg two--leg ladder as we have seen above \cite{RVA1} and to two
generalizations of it, the t-J \cite{RVA2} and the Hubbard \cite{RVA5}
two--leg ladders where the t-J and the Hubbard Hamiltonian, respectively,
are defined on the lattice represented in the figure (\ref{ladder});
it has been also applied to other ladders, the four--leg ladder
\cite{RVA4} and the diagonal ladders \cite{RVA3}, and to the spin--1
chain \cite{RVA6}. In this latter case, very accurate calculations have
been made by considering several local configurations (instead of our
two retained here); the results are compared with Density Matrix
Renormalization Group (DMRG) calculations \cite{white-huse}. We see
that, in any case, the
comparison are very good and even sometimes the RVA method gives
better results than the DMRG (see Table II). A more complete review about RVA can be found in the
following reference \cite{dmrg}.

\section{RVA method applied to Conjugated Polymers}

The low--energy properties of conjugated polymers involve
$\pi$--electrons. Usually, they are described by the use of effective
models as the well--known Pariser--Parr--Pople Hamiltonian
\cite{PP,P}. This model includes both electron--phonon interaction
(semi--classically) and a long range Coulomb interaction. The
characteristic values for each kind of energetic contribution,
kinetic, electron--phonon and electron--electron terms are
approximatively of the same order of magnitude \cite{revue}; in this
regime, the study of the PPP Hamiltonian -- or some other short
versions of it -- are really not obvious.

Compared to the previous model used in section (2), the spin 1/2
antiferromagnetic Heisenberg model for a two--leg ladder, the PPP -- or
related -- model is much more complicated: the long--range part of the
Coulomb interaction is very difficult to treat and the dynamical
variable are still not only spin but electron which involves spin and
charge degree of freedom. However, due to the alternance between
monomers, which can be large, and the link (the single bond) between
the monomers, we will underline in this work a remarkable similarity
between conjugated polymers and two--leg ladder. A similar wave function than
(\ref{18}) are proposed here for conjugated polymers - the rung of the
ladder is played by the monomer and the resonon excitation of the
two--leg ladder by a set of intermonomer
fluctuations \cite{pleutin1,pleutin2}. We believe this wave function
a natural and good starting point for further refinements.

\begin{figure}[h]
\centerline{\epsfig{file=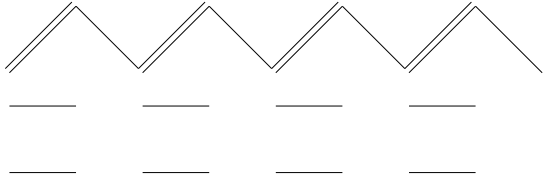,width=17pc}}
\caption[]{Model for the trans--polyacetylene and the system of bonding
-- antibonding levels associated with each double--bond}
\label{fig55}
\end{figure}

We choose as model system the simplest conjugated polymers -- the
trans--polyacetylene. This
compound shows a dimerized structure with an
alternance between double bond (1.35\AA) and single bond
(1.45\AA); the monomer is then simply a double bond (see figure (\ref{fig55})). The $\pi$
electrons are assumed to be effectively describe not by the full PPP
Hamiltonian but
by the Extended Peierls--Hubbard model (EPH) \cite{revue,eric}, a short version of it, for
simplicity

\begin{equation}
H=-\sum_{n,\sigma}(t-(-1)^{n}\frac{\Delta}{2})(c^{\dagger}_{n+1,\sigma}c^{\quad}_{n,\sigma}+c^{\dagger}_{n,\sigma}c^{\quad}_{n+1,\sigma})+U\sum_{n}n_{n
\uparrow}n_{n \downarrow}+V\sum_{n}n_{n}n_{n+1}
\end{equation}The operator $c^{\dagger}_{n,\sigma}$
($c^{\quad}_{n,\sigma}$) creates (annihilates) an electron of spin
$\sigma$ at site $n$, $n_{n
\sigma}=c^{\dagger}_{n,\sigma}c^{\quad}_{n,\sigma}$ and $n_{n}=n_{n
\uparrow}+ n_{n \downarrow}$. $t$ is the nearest--neighbour hopping
term without dimerization, $\Delta$ is the usual dimerization order
parameter, $U$ is the on-site Hubbard repulsion and $V$ is the
nearest--neighbour charge density -- charge density interaction. We
will note in the following for convenience, $t_{d}=t+\frac{\Delta}{2}$
and $t_{s}=t-\frac{\Delta}{2}$, the hopping term for the double and
the single bond respectively.

Because of the dimerization, the one electron density is more
pronounced on the double bonds than on the single bonds; this is more
and more apparent when the dimerization increases. More generally, this is the
case for every conjugated polymers where the one--electron density is
peaked on the monomer region. In this sense, conjugated polymers are
not strictly one dimensional systems but rather intermediate between
quasi--zero and quasi--one dimensional systems \cite{pleutin1,soos,mukho}. It seems
then convenient to perform a unitary transformation which favours the orbitals
localized on the monomers, in our case, on the double bonds. The new operators are then given by

\begin{equation}
B^{(\dag)}_{n,\sigma}=\frac{1}{\sqrt{2}}(c^{(\dag)}_{2n,\sigma}+
c^{(\dag)}_{2n+1,\sigma})\quad ,\quad A^{(\dag)}_{n,\sigma}=\frac{1}{\sqrt{2}}
(c^{(\dag)}_{2n,\sigma}-c^{(\dag)}_{2n+1,\sigma})
\label{dimer}
\end{equation}These operators create or annihilate $\pi$ electrons in
the corresponding bonding or antibonding orbitals of the double bond
$n$ (see figure (\ref{fig55})).

As for the case of the two--leg ladder, we select the Local
Configurations (LC) which are the much important for the ground state wave
function. This selection of the relevant LC is made by combining
energetic and symmetry considerations. For that purpose, we use the electron--hole symmetry
operator, $\hat{J}$, to classify the LC. On the Fock
space of a single site $n$, the action of this operator are summarized as
follows \cite{ramasesha}

\begin{equation}
\hat{J}_{n}\mid 0 > = \mid \uparrow \downarrow >, \quad \hat{J}_{n}\mid \sigma > =
(-1)^{n} \mid \sigma >, \quad \hat{J}_{n}\mid \uparrow \downarrow > =- \mid 0 >
\end{equation}The electron--hole symmetry operators for several site is
just given by the direct product of such single site operator
$\hat{J}=\prod_{n}\hat{J}_{n}$.

\begin{figure}[h]
\centerline{\epsfig{file=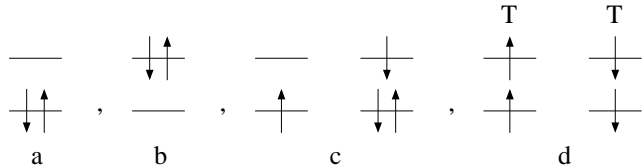,width=20pc}}
\caption[]{The set of local configurations considered in this work for
the dimerized chain; a) the $F$-LC b) the $D$-LC c) the $Ct_{1}$-LC
and d) the $TT$-LC made by combining two localized triplet symbolized
by the T on the figure.}
\label{lc}
\end{figure}

The electronic configurations can be
splitted into two classes of electron--hole symmetry denoted by (+) and (--). The
ground state is in the (+) sector,
$\hat{J}\mid GS>=\mid GS>$;
therefore we select the lowest LC in energy which belong to this
class of symmetry. We kept only four of such LC, represented in
figure (\ref{lc}) which were shown as the most important ones in finite
cluster diagonalizations \cite{chandross}. We present briefly now these
LC; their associated creation operator and their energy. They are
classified in two categories following their extension: the LC
localized on one double bond are called Molecular Configurations (MC)
-- to underline the analogy with molecular crystals; the LC extended over two
nearest--neighbour double--bonds are named
Nearest--Neighbour--Intermonomer--Fluctuations (NNIF). In such
classification, the singlet on the rungs of the previous section
enters in the first class and the resonon in the second one.

\subsection{Molecular Configurations (MC)}
The MC are the LC completely localized on the monomers. They give an
appropriated description of the system in the strong dimerized limit. There
is two kinds of MC in the (+) electron--hole symmetry sector \cite{pleutin1,pleutin2}.
\begin{itemize}
\item In the first, the monomer $n$ is in its ground state; it is described
by the following creation operator

\begin{equation}
F^{\dag}_{n}=B^{\dag}_{n,\uparrow}B^{\dag}_{n,\downarrow}
\end{equation}For a reasonable choice of parameters, this is the
lowest LC in energy so that we choose as reference state
\begin{equation}
\mid 0>= \prod_{n}F^{\dag}_{n}\mid Vacuum>
\end{equation}where $\mid Vacuum>$ is the state without any $\pi$
electron. With respect to this reference state, the local operator $F^{\dag}_{n}$
is simply given by the Unity $I_{n}$.
We have named this LC, $F$-LC. In the following all the creation
operators and the energies are defined with respect to $\mid 0>$.
\item In the second, the monomer $n$ is doubly excited; this LC is
associated with the following local operator

\begin{equation}
D^{\dag}_{n}=A^{\dag}_{n,\uparrow}A^{\dag}_{n,\downarrow}
B_{n,\uparrow}B_{n,\downarrow}
\end{equation}with energy given by $\epsilon_{d}=4t_{d}$. This LC
introduces electronic correlation inside a particular double bond. We have
named this LC $D$-LC.
\end{itemize}

Notes that the LC where a monoexcitation take place in a double
bond, let's call it $S$-LC, lower in energy than the $D$-LC, is in the
(--) electron--hole symmetry sector; therefore, it will be a local
constituent for excited states \cite{pleutin1,pleutin2}.

These two LC are the analogue of the singlet along the rung for the
two--leg ladder.

\subsection{Nearest--Neighbour--Intermonomer--Fluctuations (NNIF)}
With the two MC, the electrons are coupled by pairs on the double
bonds and the resulting picture is the one of a
molecular crystal in its ground state \cite{pleutin1,simpson,mukho};
the analogue for the two--leg ladder is given by the state $\mid 0>$
represented in figure (\ref{0resonon}) with only singlets along the rungs. This crude
picture is obviously not
adapted for conjugated polymers where $\pi$ electrons are delocalized
on the whole chain. As in the case of the two--leg ladder, fluctuations must be
included in our treatment, we talk in this case of intermonomer fluctuations. This will be done in a somehow 'minimal'
way by considering the two following LC, extending over two double
bonds only \cite{pleutin1,pleutin2}.
\begin{itemize}
\item In the first, one electron is transfered from one monomer to the
nearest--neighbour one; this operation is associated with the local
operator written as
\begin{equation}
Ct^{\dag}_{n}=\frac{1}{2}(A^{\dag}_{n+1,\uparrow}
B_{n,\uparrow}+A^{\dag}_{n+1,\downarrow}B_{n,\downarrow}-
A^{\dag}_{n,\uparrow}B_{n+1,\uparrow}-A^{\dag}_{n,\downarrow}B_{n+1,\downarrow})
\end{equation}with energy $\epsilon_{ct}=2t_{d}+\frac{3}{4}V$. This LC introduces some nearest--neighbour intermonomer
charge fluctuation reproducing in a minimal way the conjugaison phenomenom. We named this LC, $Ct_{1}$-LC.

\item In the second, some intermonomer spin fluctuation is introduced by
combining two n.n. triplets into a singlet; the corresponding operator is
\begin{equation}
\begin{array}{c}
TT^{\dag}_{n}=\frac{1}{\sqrt{3}}(A^{\dag}_{n,\uparrow}B_{n,\downarrow}A^{\dag}_{n+1,\downarrow}B_{n+1,\uparrow}+A^{\dag}_{n,\downarrow}B_{n,\uparrow}A^{\dag}_{n+1,\uparrow}B_{n+1,\downarrow}+\\
\frac{1}{2}(A^{\dag}_{n,\uparrow}B_{n,\uparrow}A^{\dag}_{n+1,\uparrow}B_{n+1,\uparrow}+A^{\dag}_{n,\uparrow}B_{n,\uparrow}A^{\dag}_{n+1,\downarrow}B_{n+1,\downarrow}+\\A^{\dag}_{n,\downarrow}B_{n,\downarrow}A^{\dag}_{n+1,\uparrow}B_{n+1,\uparrow}+A^{\dag}_{n,\downarrow}B_{n,\downarrow}A^{\dag}_{n+1,\downarrow}B_{n+1,\downarrow}))
\end{array}
\end{equation}with energy given by $\epsilon_{tt}=4t_{d}-(U-V)$; this
LC is named $TT$-LC.

\end{itemize}

There is more LC extended over two n.n. double--bonds within the (+)
electron--hole symmetry class -- and, of course, there is much more LC
more extended -- but we believe the two selected ones are the more
important. For instance, we saw that the $S$-LC is in the (--) class
of symmetry, however, with the simple product form of the
electron-hole symmetry operator, a LC with two $S$-LC -- or more
generally with an even number of $S$-LC -- is in the (+) class of
symmetry and
should be considered here. However, this kind of LC are sufficiently
high in energy to be reasonably neglected.

These two LC are the analogue of the resonon configuration for the
two--leg ladder.

\subsection{Recurrence relation for the ground state wave function}

With our choice of four LC, all possible electronic configurations
are then build up. They are characterized by the number of $D$, $Ct_{1}$ and
$TT$-LCs, $n_{d}$, $n_{ct}$ and $n_{tt}$ respectively, and by the
positions of these different local configurations. The positions of the $D$, $Ct_{1}$ and $TT$-LCs
are labelled by the coordinates $z(k)$ ($k=1,..,n_{d}$), $y(j)$
($j=1,..,n_{ct}$) and $x(i)$ ($i=1,..,n_{tt}$) respectively;
implicitly, the necessary non--overlapping condition between LC is supposed to
be fulfilled all along the paper. The electronic configurations are then expressed as

\begin{equation}
\label{espacemodel}
\begin{array}{c}
\mid x(1),...,x(n_{tt}),y(1),...,y(n_{ct}),z(1),...,z(n_{d})>=\\
\\
\prod_{i=1}^{n_{tt}}\prod_{j=1}^{n_{ct}}\prod_{k=1}^{n_{d}}TT^{\dag}_{x(i)}Ct^{\dag}_{y(j)}D^{\dag}_{z(k)}\mid
0>
\end{array}
\end{equation}We look for a ground state of the following form

\begin{equation}
\label{gs}
\mid GS
>=\sum_{\{n_{d},n_{ct},n_{tt}\}}u_{f}^{N-n_{d}-2(n_{ct}+n_{tt})}u_{d}^{n_{d}}u_{IF}^{n_{ct}+n_{tt}}u_{ct}^{n_{ct}}u_{tt}^{n_{tt}}\prod_{k=1}^{n_{d}}TT^{\dag}_{x(i)}Ct^{\dag}_{y(j)}D^{\dag}_{z(k)}\mid
0>
\end{equation}where $u_{f}$, $u_{d}$, $u_{IF}$, $u_{ct}$ and $u_{tt}$
are parameters to be determined and where the summation runs over
all the possible configurations.

This state is a kind of generalization of the Dimer--RVB state of the
two--leg ladder. The singlet on the rung is replaced by a two component
MC, which contain $F$-LC and $D$-LC, and the resonon is replaced by a two
component NNIF, which contains $Ct_{1}$-LC and $TT$-LC. By analogy with
the previous case, we can write a recurrent relation followed by the
wave function (\ref{gs}); it is given by

\begin{equation}
\label{ansatz}
\begin{array}{c}
|N>=u_{f}|N-1>\otimes|F_{N}>+u_{d}|N-1>\otimes|D_{N}>+\\
\\u_{IF}(u_{ct}|N-2>\otimes|Ct_{N-1,N}>+u_{tt}|N-2>\otimes|TT_{N-1,N}>)
\end{array}
\end{equation}where $|F_{N}>$ and $|D_{N}>$ are the MC created by
$F^{\dag}_{N}$ and $D^{\dag}_{N}$ respectively and $|Ct_{N-1,N}>$ and
$|TT_{N-1,N}>$ are the NNIF created by the $Ct^{\dag}_{N-1}$ and
$TT^{\dag}_{N-1}$ respectively. There are two constraints due to
normalization to be fulfilled by the coefficients, explicitely
$u_{f}^{2}+u_{d}^{2}=1$ and $u_{ct}^{2}+u_{tt}^{2}=1$. Among the five
parameters, three remain free and have to be determined variationally by
minimalization of the total energy per unit cell.

\subsection{Recurrence relations for the energy}

The energy per unit cell for a polymer is given again by
\begin{equation}
\epsilon_{\infty}=\lim_{N\rightarrow \infty}\epsilon_{N}=\lim_{N\rightarrow \infty}\frac{1}{N}\frac{E_{N}}{Z_{N}}
\end{equation}where
\begin{equation}
E_{N}=<N|H|N> \quad and \quad Z_{N}=<N|N>
\end{equation} These two quantities, the expectation value and the
norm of the wave function, are solutions of a set of coupled recurrent
relations, as for the two--leg ladder case. In this case, they are
simpler because the selected LC, MC and NNIF, are orthogonal; we don't
have then to consider quantities as $D_{N}$ and $Y_{N}$ and, finally, we
get only two coupled recurrent relations instead of four in the
previous section

\begin{equation}
Z_{N}=Z_{N-1}+u_{IF}^{2}Z_{N-2}
\label{rrZ}
\end{equation}
\begin{equation}
E_{N}=E_{N-1}+u_{IF}^{2}E_{N-2}+\epsilon_{m}(u_{f})Z_{N-1}+W_{2}(u_{f},u_{IF},u_{ct})Z_{N-2}
\label{rrE}
\end{equation}where $\epsilon_{m}(u_{f})$ is the correlated ground
state energy of a double bond given by
\begin{equation}
\epsilon_{m}(u_{f})=-4u_{d}^{2}t_{d}+u_{f}u_{d}(U-V)
\label{EMC}
\end{equation}and $W(u_{f},u_{IF},u_{ct})$ expressed the energy of the
NNIF-LC and their coupling with the MC
\begin{equation}
W_{2}(u_{f},u_{IF},u_{ct})=u_{IF}^{2}u_{ct}^{2}\epsilon_{ct}+u_{IF}^{2}u_{tt}^{2}\epsilon_{tt}+u_{IF}^{2}u_{tt}u_{ct}\sqrt{3}t_{s}+2u_{f}^{2}u_{IF}u_{ct}t_{s}
\label{ENNIF}
\end{equation}
With the two recurrent relations, we associated the following natural initial
conditions $Z_{0}=Z_{1}=1$ and $E_{0}=0$,
$E_{1}=\epsilon_{m}(u_{F})$.

The set of equations (\ref{rrZ}) and (\ref{rrE}) together with the
definition (\ref{EMC}), (\ref{ENNIF}) and the initial
conditions aforementioned, can be solved analyticaly and are also really
natural to implement on a computer.

In the following, comparisons are made with Density Matrix
Renormalization Group results \cite{white,eric,dmrg}. 
A finite system  DMRG algorithm is used to calculate
the ground state energy for several chain lengths up to four 
hundred sites and the energy per site is then
extrapolated to the infinite lattice limit.
The numerical error in the energy per site of the infinite
lattice is estimated to be of the order of $10^{-4}$
or smaller.
We compare the RVA results with these extrapolated values.

\begin{table}
\caption{Energies per site from DMRG and RVA calculations, $e_{\infty}^{{\rm
DMRG}}$ and $e_{\infty}^{{\rm RVA}}$ respectively, for $U=3t$ and $V=1.2t$}
\begin{tabular}{r@{\hspace{1.5cm}}r@{\hspace{1.5cm}}r} 
\hline \rule[-1.5mm]{0mm}{5mm}
$\Delta$ & $e_{\infty}^{{\rm
DMRG}}/t$ & $e_{\infty}^{{\rm
RVA}}/t$ \\  
\hline

  $0.1$ & $0.330670$ & $0.396849$ \\ 

  $0.3$ & $0.284416$ & $0.327386$ \\

  $0.5$ & $0.224344$ & $0.254182$ \\ 

  $1.5$ & $-0.175762$ & $-0.164041$ \\ 
 
\hline
\end{tabular}
\label{Tab3}
\end{table}

A reasonable choice of parameters for the polyacetylene is given by
$U=3t$ and $V=1.2t$ \cite{bursill,shuai}. In table III, we show the
energy per site obtained from DMRG and RVA calculations for different
values of the dimerization parameter $\Delta$. For $\Delta=2$, the
chain is fully dimerized; the results given by the
ansatz (\ref{gs}) is then very accurate per construction. Next, the accuracy of the RVA results
decreases when the dimerization decreases; this appear clearly in the
table III. For values relevant for conjugated polymers, $\Delta$
larger than 0.25, the energy approaches 90$\%$ of the DMRG energy; we
think these results satisfactory according to the relative simplicity
of our proposed trial wave function (\ref{gs}): only four local
configurations are retained! This wave function is, in a sense, an
extrapolation of the wave function analysed in small cluster
calculations \cite{chandross}, to
an infinite lattice. Of course some improvements are
suitable and may include more extended LC; by the way, works are
currently in progress in that direction by using again the Recurrent
Variational Approach. Nevertheless, we believe
the wave function (\ref{gs}) sufficient to get a first explanation of
some physical phenomena as the very strong linear absorption due to excitonic
states observed in polydiacetylenes \cite{pleutin1}.

\begin{figure}[h]
\centerline{\epsfig{file=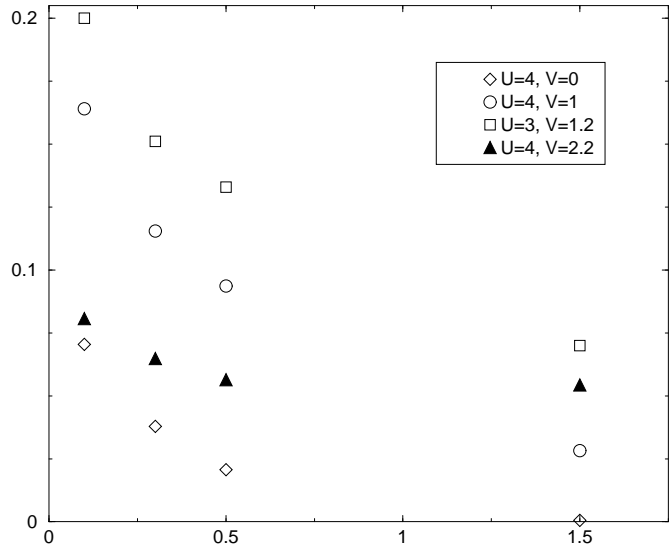,width=17pc,angle=-90}}
\caption[]{Relative errors of the RVA results compared to the DMRG
ones. The unfilled symbols are for calculations where $U>2V$; these
choices give a BOW ground state. On the contrary, the filled triangles
correspond to calculations where $U<2V$; then a CDW ground state is stabilized.}
\label{dmrgrva}
\end{figure}

Last, it is well known that the ground state of the EPH model shows a Bond Order Wave if
$U>2V$ and a Charge Density Wave in the other case \cite{mazumdar}. In
figure (\ref{dmrgrva}), we show the relative error of the RVA results
compared to the DMRG ones for several choices of Coulombic parameters
in function of the dimerization parameter $\Delta$, in the two
different regimes. Once again, the results show clearly that our ansatz
is better when $\Delta$ increases. Moreover, its behaviour seems
different in the two sides of the transition, this is natural since our
ansatz is not appropriated to describe properly a Charge Density Wave.

\section{Conclusions and Perspectives}

To conclude, we have presented in this note the Recurrent Variational
Approach and, more specifically, its first application to the
electronic structure of conjugated polymers. The ansatz built here
(\ref{gs}), shows some nice similarities with the trial state
(\ref{23}) made for the two--leg ladder.

By comparisons with Density Matrix Renormalization Group calculations,
we think this trial wave function encouraging and a good starting
point for further improvements. Moreover, we believe such kind of
simple wave functions useful to get some analytical insight into
physical phenomena \cite{pleutin1,pleutin2}.

Two natural extensions of this work are currently in progress: first, it is
possible, in some extents, to improve the simple ansatz (\ref{gs}) by
including more local configurations; second, it is also possible to
apply the method to other conjugated polymers of current interests as
the poly--paraphenylene and the poly--paraphenylenevinylene.

\end{article}
\end{document}